\documentclass[aps,prd,showpacs,nofootinbib,preprintnumbers,twocolumn]{revtex4-1}

\usepackage{bm}
\usepackage{latexsym}
\usepackage{natbib}
\usepackage{url}
\usepackage{dcolumn}
\usepackage{color}
\usepackage{amsfonts,amssymb,amsmath}
\usepackage{graphicx,epsfig}
\usepackage{psfrag}
\usepackage{subfigure}
\usepackage{natbib}

\begin{document}

\title{Dynamics of $f(R)$ gravity models and  asymmetry of time}

%--------------------------------------------------------------------------------------------------------------------------------

\author{Murli Manohar Verma}
\email[]{sunilmmv@yahoo.com}

\author{Bal Krishna Yadav}
\email[]{balkrishnalko@gmail.com}

\affiliation{Department of Physics, University of Lucknow, Lucknow 226 007, India}
%-----------------------------------------------------------------------------------------------------------------%
\date{\today}

\begin{abstract}
 We solve the field equations of modified gravity for $f(R)$ model in metric formalism. Further, we obtain the fixed points of  the dynamical system in  phase space analysis of $f(R)$  models, both with and without the effects of radiation. Stability of these points is studied against perturbations in a smooth spatial background by applying  the conditions on the eigenvalues of the matrix obtained in the linearized first-order differential equations.  Following this, these  fixed points are used for analysing  the dynamics of the system during the  radiation, matter and acceleration dominated  phases of the universe. Certain linear and quadratic forms of $f(R)$  are determined from the geometrical and physical considerations and the behaviour  of the scale factor is found for those forms. Further, we also  determine the Hubble parameter  $H(t)$, Ricci scalar $R$ for these cosmic  phases. We show the emergence of an asymmetry of time  from  the dynamics of the  scalar field  exclusively owing to the $f(R)$ gravity in the  Einstein frame that may lead to an arrow of time at a classical level.

\end{abstract}
%\keywords{}

\maketitle

\section{\label{1}Introduction}
 The present state of the universe has been  found to be in the phase of accelerated expansion\cite{a1}. There are several  observational evidences on geometry and growth of structures  such as the  Supernovae Ia, Baryon Acoustic Osicllation (BAO), Cosmic Microwave Background anisotropies, weak gravitational lensing  etc. \cite{a1,a2,a3,a4} which indicate the  presence of a  hitherto unknown  dark energy.  By all reckoning, the explanation for present accelerated expansion of the universe is a major challenge in cosmology, even though there are many  approaches to explain its dynamics. The simplest candidate for dark energy is the cosmological constant  with a constant equation of state ($w=-1$)  \cite{a5}. However,  there are two main difficulties  associated  with the cosmological constant -- (i) the fine tuning problem and (ii) the coincidence problem.  Besides,  there exist  two basic approaches attmpting to explain  dark energy. The first approach is based on modified matter models. In this approach $T_{\mu\nu}$ in the  Einstein equations must  include  an exotic matter component like quintessence, k-essence, Phantom energy etc. \cite{a6,a7,a8,a9} that takes the form  of dark energy and so  the resposibility of causing acceleration. The second approach is through the so-called  modified gravity models wherein the  late-time accelerated cosmic expansion is realized without requiring  the  explicit dark energy  component in the universe. In these  models, we have a wide range  of  $f(R)$ gravity \cite{a10}, scalar-tensor theories, Gauss-Bonnet models, braneworld models  etc. \cite{a11,a12,a13}. Specifcally, in  this paper we present an analysis of  $f(R)$  models, where  one modifies the laws of gravity by replacing  the  scalar curvature $R$ of the Hilbert's action, or $R-2\Lambda$, as  one includes in the  standard $\Lambda$CDM approach  (with  $\Lambda$  as the cosmological constant),   by an arbitrary function of  $R$ in the curvature part of the  Lagrangian density.   At present, there is no specific,  known  functional form  of $R$ which may satisfy all the observational conditions of cosmological viability ranging from the radiation dominated  matter to the ongoing  accelerated phase. Therefore,  we study  the stability conditions for the  respective eras and determine  the corresponding forms of $f(R)$. By solving the field equations for different forms of $f(R)$,  the scale factor of expansion is thus determined. From here we find the scalar curvature $R$ and compare them in different eras. This can lead to the determination of a time-ordering  of various epochs,  dominated by radiation, matter and dark energy (as a modification of gravity), respectively,  throughout the evolution of the universe.

 In section II, the fixed points of the dynamical system are  determined within the framework of $f(R)$ models in metric formalism. To study the exclusive effects of radiation, in sections  III and IV the  properties and stability of the fixed points of the dynamical system  are found  without and with radiation. Section V comprises of analysis of the  behaviour of  $f(R)$, scale factor $a(t)$, Hubble parameter $H(t)$ (under various conditions)  and scalar curvature $R$ in radiation dominated phase.   In sections VI and  VII we attempt to determine  the form of $f(R)$, scale factor $a(t)$ and scalar curvature $R$ for matter dominated and the present accelerated expansion dominated phases, respectively. Together, the time ordering  may be used further to determine an arrow of time through the cosmic evolution in section VIII.
Finally,  we conclude our results in section IX.

\section{\label{2}Field equations and phase space dynamics }
In $f(R)$  gravity we obtain the  field equations in metric formalism, where  the  variation of the action is taken with  respect to $g_{\mu\nu}$ related to the connections $\Gamma^\alpha_{\beta\gamma}$ in the usual sense (unlike the Palatini formalism  where they are treated as mutually  independent).
We consider the field equations in the background of spatially flat Friedmann-Lemaitre-Robertson-Walker (FLRW) spacetime with a metric
\begin{eqnarray}ds^{2}= -dt^{2} + a^2(t)[dr^2 + r^2 (d\theta^{2} + \sin^{2}\theta d\phi^{2})]\label{a0} \end{eqnarray}
where $a(t)$ is time dependent scale factor and the speed of light $c=1$.
Correspondingly, the  Ricci scalar $R$ is given by
\begin{eqnarray}R = 6(2H^2 + \dot{H})\label{a1} \end{eqnarray}
where $H (=\dot{a}/a) $ is the Hubble parameter and an overdot represents the  derivative with respect to  time.
The total action  using an arbitrary function  $f(R)$  in the Jordan frame is given by
\begin{eqnarray}\mathcal{A} = \frac{1}{2\kappa^{2}}\int\sqrt{-g}f(R)d^{4}x + \mathcal{A}_{m} \label{a2}\end{eqnarray}
where $\mathcal{A}_{m}$ is the action for relativistic and non-relativistic matter, $\kappa^{2}= 8\pi G$ and $g$ is the determinant of the metric tensor $g_{\mu\nu}$.
Varying the action (\ref{a2}) with respect to  $g_{\mu\nu}$, the field equations obtained are
\begin{eqnarray}F(R)R_{\mu\nu} - \frac{1}{2} f(R)g_{\mu\nu}\nonumber\\ - \nabla_\mu\nabla\nu F(R) + g_{\mu\nu}\Box F(R) = \kappa^{2}T_{\mu\nu},\label{a3}\end{eqnarray}
where $F(R)\equiv\frac{\partial{f}}{\partial{R}}$ and $T_{\mu\nu}$ is the  energy-momentum tensor for matter.
From the above equation (\ref{a3}) and its trace  we arrive at the  the following ones
\begin{eqnarray}3FH^{2}= \kappa^{2}(\rho_{m}+\rho_{r})+\frac{(FR-f)}{2}-3H\dot{F}\label{a4}\end{eqnarray}
\begin{eqnarray}-2F\dot{H}=\kappa^{2}(\rho_{m}+\frac{4}{3}\rho_{r})+\ddot{F}-H\dot{F}\label{a5}\end{eqnarray}
where $\rho_{m}$  and  $ \rho_{r}$ are the  energy densities of matter and radiation,  respectively.

With  four (dimensionless)  variables  defined as
\begin{eqnarray} x_{1}\equiv - \frac{\dot{F}}{F H}\label{a7}\end{eqnarray}
\begin{eqnarray}x_{2} \equiv - \frac{f}{6F H^{2}}\label{a8} \end{eqnarray}
\begin{eqnarray}x_{3}\equiv \frac{R}{6H^{2}}\label{a9} \end{eqnarray}
\begin{eqnarray}x_{4} \equiv \frac{\kappa^{2}\rho_{r}}{3 F H^{2}} \label{a10} \end{eqnarray}
the effective equation of state for this  system is defined by
\begin{eqnarray}w_{eff}= -1 -\frac{2}{3}\frac{\dot{H}}{H^2} = -\frac{1}{3}(2x_{3} - 1)  \label{a6}\end{eqnarray}

Differentiation of these variables (\ref{a7}-\ref{a10}) with respect to   $N=\ln a(t)$  gives
\begin{eqnarray} \frac{dx_{1}}{dN} = -1 -x_{3} - 3x_{2} + x_{1}^{2} - x_{1} x_{3} + x_{4}\label{b1}\end{eqnarray}

\begin{eqnarray} \frac{dx_{2}}{dN} = \frac{x_{1}x_{3}}{m} - x_{2}(2x_{3} - 4 - x_{1})\label{b2}\end{eqnarray}

\begin{eqnarray} \frac{dx_{3}}{dN} = -\frac{x_{1}x_{3}}{m} - 2x_{3}(x_{3} - 2)\label{b3}\end{eqnarray}

\begin{eqnarray} \frac{dx_{4}}{dN} = -2x_{3} x_{4} + x_{1} x_{4}\label{b4}\end{eqnarray}
where
\begin{eqnarray}m\equiv\frac{d\log F}{d\log R}= \frac{Rf_{,RR}}{f_{,R}},\label{b5}\end{eqnarray}
\begin{eqnarray}q\equiv - \frac{d\log f}{d \log R}=-\frac{Rf_{,R}}{f} = \frac{x_{3}}{x_{2}}\label{b6}\end{eqnarray}
where $f_{,R}\equiv \frac{df}{dR}$ and $f_{,RR}\equiv \frac{d^2 f}{dR^2}$.
The fixed points of the system   are obtained by equating the equations  (\ref{b1} - \ref{b4})  to zero.  Thus, the   points are given by
\begin{eqnarray} P_{1}:(x_{1}, x_{2}, x_{3}, x_{4}) = (0, -1, 2, 0),\nonumber\\ \Omega_{m}=0, w_{eff}= -1\label{b7} \end{eqnarray}
\begin{eqnarray}P_{2}:(x_{1}, x_{2}, x_{3}, x_{4}) = (-1, 0, 0, 0), \nonumber\\\Omega_{m}=2, w_{eff}= \frac{1}{3}\label{b8}  \end{eqnarray}
\begin{eqnarray}P_{3}:(x_{1}, x_{2}, x_{3}, x_{4}) = (1, 0, 0, 0),\nonumber\\\Omega_{m}=0,  w_{eff}=\frac{1}{3} \label{b9} \end{eqnarray}
\begin{eqnarray} P_{4}:(x_{1}, x_{2}, x_{3}, x_{4}) = (-4, 5, 0, 0),\nonumber\\ \Omega_{m}=0,  w_{eff}= \frac{1}{3}, \label{b10}\end{eqnarray}
\begin{eqnarray}P_{5}:(x_{1}, x_{2}, x_{3}, x_{4}) = \nonumber\\ (\frac{3m}{1+m},-\frac{1+4m}{2(1+m)^{2}}, \frac{1+4m}{2(1+m)}, 0 ),\nonumber\\ \Omega_{m}= 1 - \frac{m(7 + 10m)}{2(1+m)^{2}}, \nonumber\\w_{eff}= -\frac{m}{(1+m)}, \label{c1}\end{eqnarray}
\begin{eqnarray} P_{6}:(x_{1}, x_{2}, x_{3}, x_{4}) = \nonumber\\ \left(\frac{2(1-m)}{1+2m}, \frac{1- 4m}{m(1+2m)}, -\frac{(1-4m)(1+m)}{m(1+2m)}, 0 \right),\nonumber\\\Omega_{m}= 0, \nonumber\\ w_{eff}= \frac{2 - 5m - 6m^{2}}{3m(1+2m)}\label{c2}\end{eqnarray}
\begin{eqnarray}P_{7}:(x_{1}, x_{2}, x_{3}, x_{4}) = (0, 0, 0, 1) \nonumber\\  \Omega_{m}=0, \quad   w_{eff}= \frac{1}{3} \label{c3}\end{eqnarray}
\begin{eqnarray}P_{8}:(x_{1}, x_{2}, x_{3}, x_{4}) =\nonumber\\ \left(\frac{4m}{1+m},  -\frac{2m}{(1+m)^{2}}, \frac{2m}{1+m}, \frac{1- 2m -5m^{2}}{(1+m)^{2}}\right),\nonumber\\
\Omega_{m}= 0,\quad  w_{eff}= \frac{1 - 3m}{3 + 3m}\label{c4}\end{eqnarray}.
%------------------------------------------------------------------------------------
\section{\label{3}Fixed points without radiation}

 First, we  consider the properties and stability of these fixed points in the absence of radiation. For stability about the fixed points $(x_{1}, x_{2}, x_{3})$ we invoke time dependent  linear perturbations $\delta x_{i}  (i=1, 2, 3)$ around the points in a smooth spatial background.  Linearization of  the equations (\ref{b1}-\ref{b3})  gives first order differential equations
 \begin{eqnarray} \frac{d}{dN}\left(\begin{matrix}\delta x_{1}\\ \delta x_{2} \\ \delta x_{3} \end{matrix}\right)= M \left(\begin{matrix}\delta x_{1}\\ \delta x_{2} \\ \delta x_{3} \end{matrix}\right),\label{n1}  \end{eqnarray}
 where $M$ is a $3\times3$ matrix whose components depend  upon $x_{1}, x_{2}$ and
 $ x_{3}$. Stability of each fixed point depends upon the eigenvalues of the matrix $M$ obtained by taking linear  perturbations around that specific  point. In the absence of radiation, we have only six fixed points $P_{1}-P_{6}$ as below.
\begin{enumerate}
\item[(1)]  Point $P_{1}:(0, -1, 2)$ corresponds to de-Sitter point. Here $w_{eff} = -1$ and eigenvalues corresponding to this point are
\begin{eqnarray} -3,   -\frac{3}{2} \pm \frac{\sqrt{25 - \frac{16}{m}}}{2}\label{c5}\end{eqnarray}
$P_{1}$  is stable when real parts  of all the eigenvalues is negative. Hence condition for stability is $0<m(q=-2)<1$,   otherwise it is a saddle point. So this point can be taken as an acceleration point.

\item [(2)] Point $P_{2}:(-1, 0, 0)$ is denoted by $\phi$-matter-dominated ($\phi$ MDE) epoch. The eigenvalues of the $3\times3$ matrix of perturbations about $P_{2}$ are given by
\begin{eqnarray} -2,\frac{1}{2}[7+ \frac{1}{m} -\frac{m'}{m^2}q(1+q) \mp \nonumber\\ \sqrt{\left(7+ \frac{1}{m}-\frac{m'}{m^2}q(1+q)\right)^2-4\left(12+ \frac{3}{m}-\frac{m'}{m^2}q(3+4q)\right)}], \nonumber  \\ \label{n2} \end{eqnarray}
where $m'$ is derivative of $m$ w.r.t. $q$. If $m$ is constant, then eigenvalues are  $-2, 3, 4+ \frac{1}{m}$. In this case $P_{2}$ is a saddle point because eigenvalues are negative and positive.

$P_{2}$ can not be a matter dominated point because $\Omega_{m} = 2$ and $w_{eff} = \frac{1}{3}$.

\item [(3)] Point $P_{3}:(1, 0, 0)$ is the  kinetic point. The eigenvalues corresponding to this point are \begin{eqnarray} 2,\frac{1}{2}[9+ \frac{1}{m} -\frac{m'}{m^2}q(1+q) \mp \nonumber\\\sqrt{\left(9- \frac{1}{m}+\frac{m'}{m^2}q(1+q)\right)^2-4\left(20- \frac{5}{m}-\frac{m'}{m^2}q(5+4q)\right)}],\nonumber \\  \label{n3}. \end{eqnarray} If $m$ is constant, the eigenvalues are $2, 5, 4-\frac{1}{m}$.
In this case $P_{3}$ is unstable for $m<0$ and $m>\frac{1}{4}$  and a saddle otherwise.

\item [(4)] Point $P_{4}:(-4, 5, 0)$ has eigenvalues:
\begin{eqnarray}-5,    -3,   4(1 + \frac{1}{m})\label{c6}   \end{eqnarray}
It is stable for $-1< m <0$ and saddle otherwise. This point cannot be used   as a radiation or a matter dominated point.

\item [(5)] Point $P_{5}:(\frac{3m}{1+m},-\frac{1+4m}{2(1+m)^{2}}, \frac{1+4m}{2(1+m)} )$ can be regarded as a standard matter point in the limit $m\rightarrow0$. In this limit $\Omega_{m}=1$ and $a \propto t^{\frac{2}{3}}$. Hence necessary condition for this point to be a standard matter point is
\begin{eqnarray}m(q=-1)= 0. \label{n4} \end{eqnarray}
Eigenvalues corresponding to  point $P_{5}$ are given by
\begin{eqnarray} 3(1 + m'), \nonumber\\ \frac{-3m \pm \sqrt{m (256m^3 + 160m^2 - 31m -16)}}{4m(m+1)}\label{c7} \end{eqnarray}
For a cosmologically viable trajectory, we want a saddle matter point. Hence,  the condition for a saddle matter epoch is given by

\begin{eqnarray}m (q\leq -1)>0, m'(q \leq -1)> -1,\nonumber\\ m(q = -1)=0\label{c8} \end{eqnarray}

\item[(6)] Point $P_{6}:\left(\frac{2(1-m)}{1+2m}, \frac{1- 4m}{m(1+2m)}, -\frac{(1-4m)(1+m)}{m(1+2m)} \right)$ can also be an acceleration dominated point. The eigenvalues corresponding to this point are:

\begin{eqnarray} -4 + \frac{1}{m}, \frac{2-3m-8m^2}{m(1+2m)}, -\frac{2(m^2 - 1)(1+ m')}{m (1+2m)}\label{c9}\end{eqnarray}
Stability of this point depends on both $m$ and $m'$.  The condition of acceleration $(w_{eff} < -\frac{1}{3})$ depends on the value of $m$.
\end{enumerate}
\section{\label{4}Fixed points with radiation}
%-------------------------------------------------------------------------------------------

Next, we include  the radiation with other components of universe as a realistic  case for our further study. In this case we have eight fixed points. Stability about the fixed points $(x_{1},x_{2},x_{3},x_{4})$  is determined in the same way  as in absence  of   radiation. Here,  we have $4\times4$  matrix of linear perturbations about each fixed point and four eigenvalues.
\begin{itemize}
\item[(1)] Point $P_{1}$ corresponds to de-Sitter point. Here $w_{eff} =-1$ and eigenvalues corresponding to this point are
\begin{eqnarray}-4, -3,   -\frac{3}{2} \pm \frac{\sqrt{25 - \frac{16}{m}}}{2}\label{c5}\end{eqnarray}
In the presence of radiation, we have an  eigenvalue $-4$  in addition to those  in the absence of radiation. Since this eigenvalue is negative, therefore the condition of stability is the  same in both cases.
$P_{1}$  is stable when $0<m(q=-2)<1$. This point may be taken as an acceleration point. The condition of stability for  this point is same as in the case of without radiation because here we have only an extra eigenvalue $-4$, which is negative.

\item[(2)] Point $P_{2}$ is denoted by $\phi$-matter-dominated ($\phi$ MDE) epoch. The eigenvalues corresponding to this point are given by

\begin{eqnarray} -2,-1, \frac{1}{2}[7+ \frac{1}{m} -\frac{m'}{m^2}q(1+q) \mp \nonumber\\\sqrt{\left(7+ \frac{1}{m}-\frac{m'}{m^2}q(1+q)\right)^2-4\left(12+ \frac{3}{m}-\frac{m'}{m^2}q(3+4q)\right)}], \nonumber \\ \label{n5} \end{eqnarray}

 $P_{2}$  is either saddle or stable point. In this case $P_{2}$ can not be a matter point because $\Omega_{m} = 2$ and $w_{eff} = \frac{1}{3}$.

\item[(3)] Point $P_{3}$ is known as kinetic point. The eigenvalues for the $4\times4$ matrix of perturbations about this point are

\begin{eqnarray}1, 2,\frac{1}{2}[9+ \frac{1}{m} -\frac{m'}{m^2}q(1+q) \mp \nonumber\\\sqrt{\left(9- \frac{1}{m}+\frac{m'}{m^2}q(1+q)\right)^2-4\left(20- \frac{5}{m}-\frac{m'}{m^2}q(5+4q)\right)}], \nonumber \\ \label{n3}. \end{eqnarray}

If $m$ is constant, the eigenvalues corresponding to this point are $2, 5, 4-\frac{1}{m}$. In this case $P_{3}$ is unstable for $m<0$ and $m>\frac{1}{4}$  and a saddle otherwise.

\item[(4)] Point $P_{4}$ has eigenvalues
\begin{eqnarray}-5,  -4,  -3,   4(1 + \frac{1}{m})\label{c6}   \end{eqnarray}
It is stable for $-1< m <0$ and saddle otherwise. This point can not be use as a radiation or a matter dominated point.

\item[(5)] Point $P_{5}$ can be regarded as a standard matter point in the limit $m\rightarrow0$. Eigenvalues for point $P_{5}$ are given by
\begin{eqnarray} -1,3(1 + m'),\nonumber\\ \frac{-3m \pm \sqrt{m (256m^3 + 160m^2 - 31m -16)}}{4m(m+1)}\label{c7} \end{eqnarray}
where  $m'$  is derivative of $m$ w.r.t. q.
For a cosmologically viable trajectory, we want a saddle matter point. The condition for a saddle matter epoch is given by

\begin{eqnarray}m (q\leq -1)>0, m'(q \leq -1)> -1,\nonumber\\ m(q = -1)=0\label{c8} \end{eqnarray}

\item[(6)] Point $P_{6}$ can also be an acceleration dominated point. The eigenvalues corresponding to this point are given by

\begin{eqnarray} -\frac{2(-1 +2m +5m^2)}{m(1+2m)}, -4 + \frac{1}{m},\nonumber\\ \frac{2-3m-8m^2}{m(1+2m)}, -\frac{2(m^2 - 1)(1+ m')}{m (1+2m)}\label{c9} \end{eqnarray}
Stability of this point depends on both $m$ and $m'$. Condition of acceleration $(w_{eff} < -\frac{1}{3})$ depends on the value of $m$.

\item[(7)] Point $P_{7}$ corresponds to a standard radiation point. The eigenvalues of $P_{7}$ for constant $m$ are  $4, 4, 1, -1$. Thus,  $P_{7}$ is  a saddle point.

\item[(8)] Point $P_{8}$ also is a radiation point. In this case dark energy is non-zero, therefore $P_{8}$ is acceptable as a radiation point. The eigenvalues of $P_{8}$ are given by
\begin{eqnarray} 1, 4(1+m') , \frac{m-1\pm\sqrt{81m^2 +30m - 15}}{2(m + 1)}. \label{c10} \end{eqnarray}
Point $P_{8}$ is a saddle point in the limit $m\rightarrow 0$. The acceptable  radiation dominated point  $P_{8}$  lies at point $(0 , -1)$ in the $(m , q)$ plane.

\end{itemize}

\section{\label{5} Dynamics of radiation dominated phase }

For radiation dominated era, phase space analysis shows that we can find a radiation point in the limit $m\rightarrow0$ at point $P_{8}$. This point lies  on the line $m=-q-1$ in the $(m,q)$ plane. Hence, the  necessary condition for this point to exist as an exact standard radiation point is given by
\begin{eqnarray}m(q=-1)\approx0.\label{c11}\end{eqnarray}
From definition of $q$ and the above  condition, the form of $f(R)$ for radiation dominated era is given by
\begin{eqnarray}f(R)=\alpha R\label{d1}\end{eqnarray}
where $\alpha$ is an integration constant.
The standard radiation point is obtained by substitution of $m\approx0$ in the radiation point of $m(q)$ curve. In this condition, the effective equation of state is
\begin{eqnarray} w_{eff}=\frac{1}{3} \label{d2}\end{eqnarray}
Using equations (\ref{a6}) and (\ref{d2}), the Hubble parameter is given by
\begin{eqnarray}H(t) = \frac{1}{(2t - c_{1})}\label{d3}\end{eqnarray}
where $c_{1}$ is an integration constant.

%-------------------------------------------------------------------------------------------
\begin{figure}[h]
\centering  \begin{center} \end{center}
\includegraphics[width=0.50\textwidth,origin=c,angle=0]{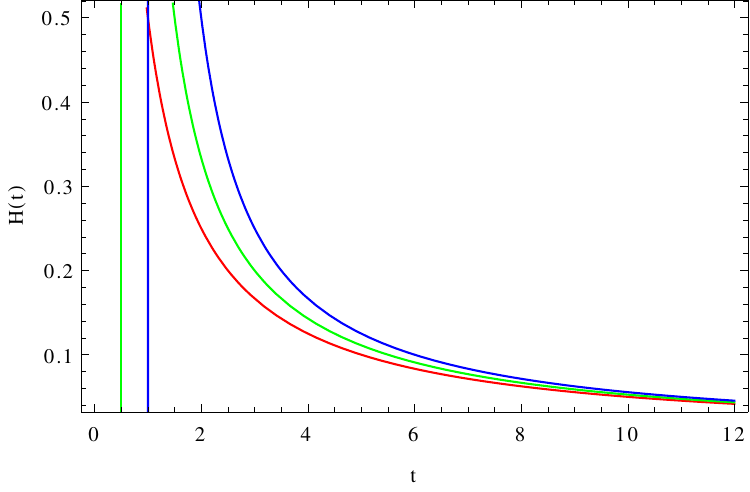}
% "\includegraphics" is very powerful; the graphicx package is already loaded
\caption{\label{fig:p1}Plot for variation of the Hubble parameter $H(t)$ with cosmic time $t$  in radiation dominated phase.  The red, green and blue   curves  correspond to $c_{1}=0,c_{1}=1,c_{1}=2$,   respectively.} \label{f1}

\end{figure}
%-------------------------------------------------------------------------------------------

The scale factor for this era is given by
\begin{eqnarray} a(t) = c_2 (2t - c_1)^\frac{1}{2} \label{d4}\end{eqnarray}
where $c_{2}$ is another integration constant.

%-----------------------------------------------------------------------------------
\begin{figure}[h]
\centering  \begin{center} \end{center}
\includegraphics[width=0.50\textwidth,origin=c,angle=0]{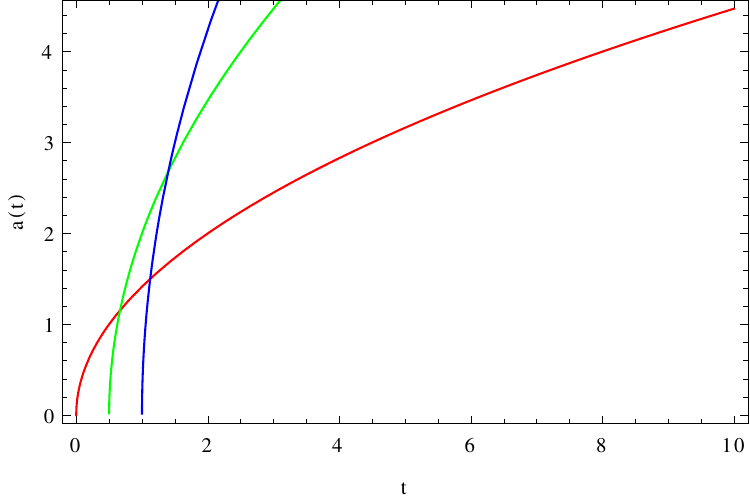}
% "\includegraphics" is very powerful; the graphicx package is already loaded
\caption{\label{fig:p2} Plot for variation of the scale factor  $a(t)$ with cosmic time  ($0\leq t \leq10$) in radiation dominated phase. The red, green and blue   curves  correspond to $(c_{1},c_{2})\equiv(0,1);(c_{1},c_{2})\equiv(1,2);(c_{1},c_{2})\equiv(2,3)$,   respectively.}\label{f2}

\end{figure}
%------------------------------------------------------------------------------------------

In radiation dominated phase we confirm  that the scale factor $a(t) \propto t^{\frac{1}{2}}$, which  is same as in the case of standard model. Figures (\ref{f1}) and (\ref{f2}) show  variation the Hubble parameter $H(t)$ and scale factor $a(t)$ with time $t$ in radiation phase.
As expected, the Ricci scalar $R$ for radiation dominated era is given by
\begin{eqnarray} R = 0\label{d5}\end{eqnarray}

\section{\label{6}Dynamics of matter dominated era }
From the field equations $(5)$ and $(6)$  we obtain the following equation
\begin{eqnarray}-\frac{\kappa^{2}\rho_{r}}{3} + 3FH^2 + F\dot{H}- \frac{f}{2} - 2H\dot{F} -\ddot{F}= 0\label{d6} \end{eqnarray}
In phase space analysis of dynamical system, there is a point $P_{5}$ which represents a standard matter era in the limit $m\rightarrow0$. In matter dominated phase of the Universe
\begin{eqnarray}   m(q=-1)\approx 0 \label{d7}          \end{eqnarray}
Using the definition of $q$ or $m$, the form of $f(R)$ is given by
\begin{eqnarray}f(R) = \beta R \label{d8} \end{eqnarray}
where $\beta$ is an  integration constant. Thus,  in matter dominated phase  the form of $f(R)$ is similar as in the case of radiation dominated phase.

In matter dominated phase, we neglect the energy density of radiation i.e.  $\rho_{r}=0$.  For $f(R) = \beta R$, $F= \beta$ and therefore  $\dot{F}=0$.  Using equations  (\ref{d6}) and (\ref{a1}) and these values of $F$ and $\dot{F}$,   the time evolution of the  Hubble parameter is expressed  as
\begin{eqnarray} H(t) = \frac{1}{(\frac{3}{2}t - c_{3})}\label{d9}\end{eqnarray}
where $c_{3}$ is an integration constant.

%----------------------------------------------------------------------------------------
\begin{figure}[h]
\centering  \begin{center} \end{center}
\includegraphics[width=0.50\textwidth,origin=c,angle=0]{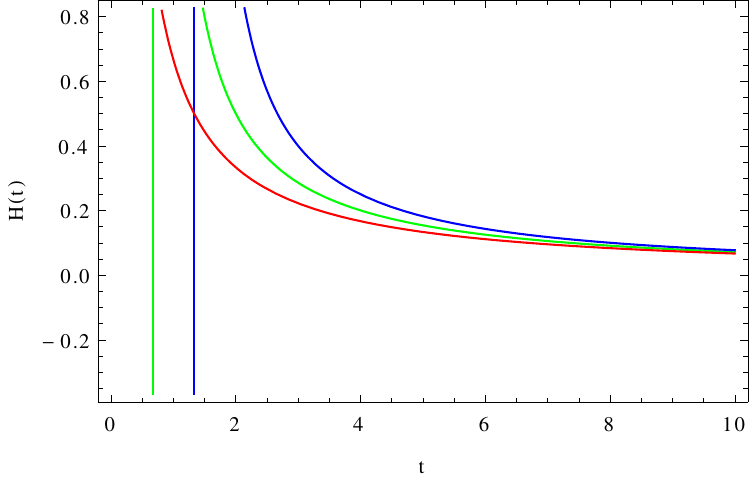}
% "\includegraphics" is very powerful; the graphicx package is already loaded
\caption{\label{fig:p3}Plot for variation of the Hubble parameter  $H(t)$ with cosmic time $t$ in matter dominated phase.The red, green and blue   curves  correspond to $c_{3}=0,c_{3}=1,c_{3}=2$,   respectively.}\label{f3}
\end{figure}
%-----------------------------------------------------------------------------------------

The scale factor in this phase is given by the expression
\begin{eqnarray}a(t) = c_{4}\left(\frac{3}{2}t - c_{3}\right)^{\frac{2}{3}}\label{d10}\end{eqnarray}

%------------------------------------------------------------------------------------------
\begin{figure}[h]
\centering  \begin{center} \end{center}
\includegraphics[width=0.50\textwidth,origin=c,angle=0]{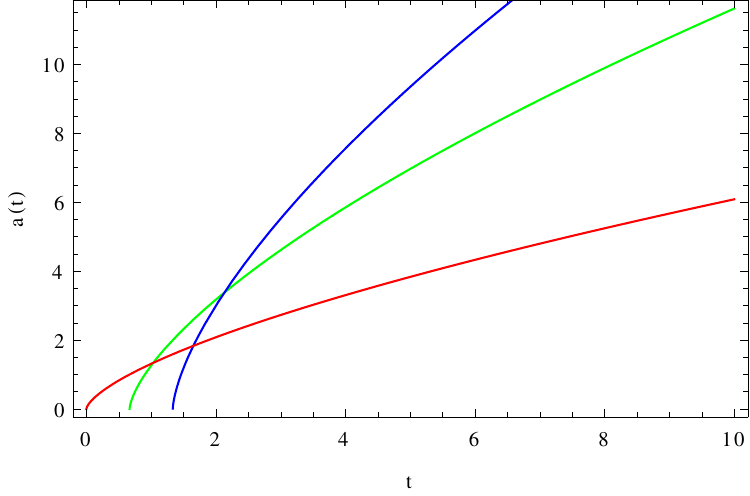}
% "\includegraphics" is very powerful; the graphicx package is already loaded
\caption{\label{fig:p4}Plot for variation of the  scale factor  $a(t)$ with cosmic time $t$  in matter dominated phase.The red, green and blue curves  correspond to $(c_{3},c_{4})\equiv(0,1);(c_{3},c_{4})\equiv(1,2);(c_{3},c_{4})\equiv(2,3)$,   respectively.}\label{f4}
\end{figure}

%----------------------------------------------------------------------------------------

 From equations (\ref{a1})  and (\ref{d9})  the Ricci scalar in matter dominated phase is given by
\begin{eqnarray} R = \frac{3}{(\frac{3}{2}t - c_3)^2}\label{e1}\end{eqnarray}

%-------------------------------------------------------------------------------------------
\begin{figure}[h]
\centering  \begin{center} \end{center}
\includegraphics[width=0.50\textwidth,origin=c,angle=0]{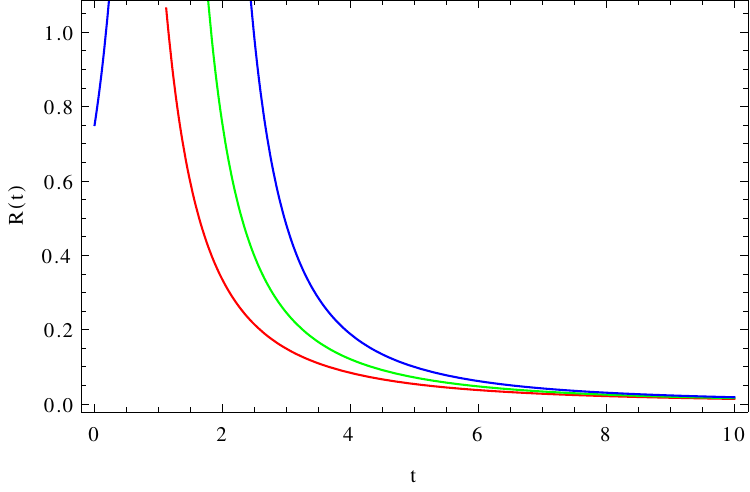}
% "\includegraphics" is very powerful; the graphicx package is already loaded
\caption{\label{fig:p5}Plot for variation of Ricci scalar $R$ with cosmic time  $t$ in matter dominated phase. The red, green and blue   curves  correspond to $(c_{3},c_{4})\equiv(0,1);(c_{3},c_{4})\equiv(1,2);(c_{3},c_{4})\equiv(2,3)$,   respectively.}\label{f5}
\end{figure}
%---------------------------------------------------------------------------------------------

The variation of Hubble parameter $H(t)$, scale factor $a(t)$  and Ricci scalar $R$,  with time is plotted in figures (\ref{f3}), (\ref{f4}), and (\ref{f5}) respectively.
Hubble parameter$H(t)$, scale factor $a(t)$, and Ricci scalar $R$ in this phase can also be calculated by the same procedure as we followed in the radiation era. Expressions for these parameters are same in both approaches.  For $m\approx 0$, the effective equation of state is given by
\begin{eqnarray}w_{eff} = 0\label{e2}\end{eqnarray}
These  expressions of scale factor $a(t)$, Hubble parameter $H(t)$, and  Ricci scalar $R$ in matter dominated phase are similar to the expressions of standard ($\Lambda$CDM) model.

\section{\label{7}Dynamics of accelerated expansion dominated phase}
In the phase space analysis, there is a point $P_{1}$,  for which effective equation of state is
\begin{eqnarray}w_{eff}= -1 \label{e3}\end{eqnarray}
This point is called de Sitter point. If we take de Sitter expansion, this point is stable when $0<m<1$ at $q=-2$. Now from the definition of $q$, the form of $f(R)$ in this phase is given by
\begin{eqnarray}f(R) = \gamma R^2\label{e4} \end{eqnarray}
We have the effective equation of state
\begin{eqnarray}w_{eff}= -1-\frac{2}{3}\frac{\dot{H}}{H^2}\label{e5} \end{eqnarray}
Now,  using equations  (\ref{e3}) and (\ref{e5}), in this phase we get the constant value of the Hubble parameter as 
\begin{eqnarray}H(t) = c_{5}\label{e6}\end{eqnarray} where $c_{5}$ is an integration constant. Therefore, the  Ricci scalar in this phase is given by
\begin{eqnarray} R = 12c_{5}^2\label{e7}\end{eqnarray}
Using the expression of Hubble parameter $H(t)$, the scale factor is given by
\begin{eqnarray}a(t) = e^{c_{5}t + c_{6}}\label{e8}\end{eqnarray} where $c_{6}$ is another integration constant.
We can also find out these parameters using equations (\ref{d6})  and (\ref{a1}) in the spatially flat universe.

%-----------------------------------------------------------------------------------
\begin{figure}[h]
\centering  \begin{center} \end{center}
\includegraphics[width=0.50\textwidth,origin=c,angle=0]{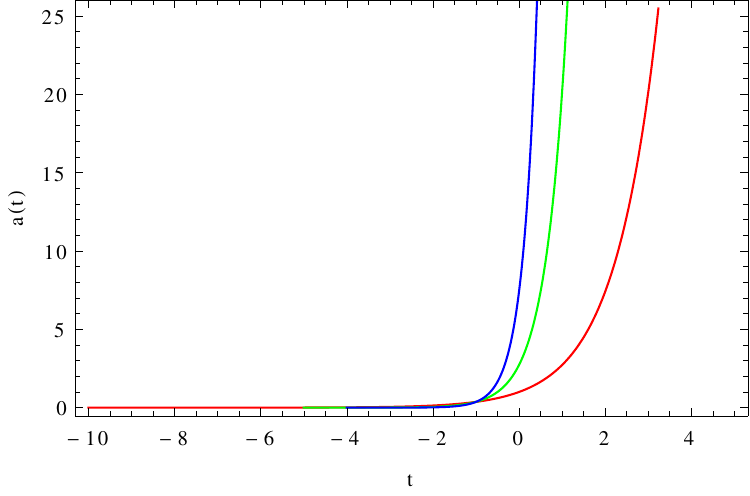}
% "\includegraphics" is very powerful; the graphicx package is already loaded
\caption{\label{fig:p6}Plot for variation of the scale factor  $a(t)$ with cosmic time $t$ in acceleration dominated phase.   The red, green and blue   curves  correspond to $(c_{5},c_{6})\equiv(1,0);(c_{5},c_{6})\equiv(2,1);(c_{5},c_{6})\equiv(3,2)$,  respectively.}\label{f6}
\end{figure}
%-----------------------------------------------------------------------------------

Here,  figure (\ref{f6}) shows the variation of scale factor $a(t)$ with time. It is clear that expansion in this phase is exponential. This behaviour is found to be  similar to the case  of the standard ($\Lambda$CDM) model.

\section{\label{8} Asymmetry  of time}
We rewrite the action (\ref{a2}) in the form
\begin{eqnarray}\mathcal{A} = \int\sqrt{-g}\left(\frac{1}{2\kappa^{2}}FR - U \right)d^{4}x + \mathcal{A}_{m},\label{g1}                                \end{eqnarray}
where \begin{eqnarray} U = \frac{FR-f}{2\kappa^2}.\label{g2} \end{eqnarray}
It is possible to derive an action in the Einstein frame under the conformal transformation

\begin{eqnarray} \tilde{g}_{\mu\nu} = \Omega^2g_{\mu\nu},\label{g3}\end{eqnarray}
where $\Omega^2$ is the conformal factor and a tilde denotes the  quantities pertaining to  the Einstein frame. The corresponding  Ricci scalars in the two frames are related  as
\begin{eqnarray} R = \Omega^2(\tilde{R} + 6\tilde{\Box}\omega - 6\tilde{g}^{\mu\nu}\partial_{\mu}\omega\partial_{\nu}\omega),   \label{g4}\end{eqnarray}
where\begin{eqnarray} \omega \equiv \ln\Omega,     \partial_{\mu}\omega\equiv\frac{\partial\omega}{\partial\tilde{x}^{\mu}},    \tilde{\Box}\omega \equiv\frac{1}{\sqrt{-\tilde{g}}}\partial_{\mu}(\sqrt{-\tilde{g}}\tilde{g}^{\mu\nu}\partial_{\nu}\omega).\label{g5}\end{eqnarray}
 Thus, the action (\ref{g1}) is transformed as \begin{eqnarray} \mathcal{A} = \int d^{4}x \sqrt{-\tilde{g}}\left[\frac{1}{2\kappa^{2}}F\Omega^{-2}(\tilde{R} + 6\tilde{\Box}\omega - 6\tilde{g}^{\mu\nu}\partial_{\mu}\omega\partial_{\nu}\omega) - \Omega^{-4}U \right] \nonumber\\ + \mathcal{A}_{m}. \nonumber \\  \label{g6}\end{eqnarray}
The linear action in $\tilde{R}$ can be written by choosing
\begin{eqnarray} \Omega^{2} = F. \label{g7}\end{eqnarray}
We  consider a new scalar field $\phi$ defined by
\begin{eqnarray} \kappa\phi \equiv \sqrt{\frac{3}{2}}\ln F. \label{g8}\end{eqnarray}

Using these relations  the action  in Einstein frame is found as \cite{a15} 
\begin{eqnarray}\mathcal{A} = \int d^{4}x \sqrt{-\tilde{g}}\left[\frac{1}{2\kappa^{2}}\tilde{R} -  \frac{1}{2}\tilde{g}^{\mu\nu}\partial_{\mu}\phi\partial_{\nu}\phi - V(\phi) \right] + \mathcal{A}_{m}. \nonumber \\  \label{g9}\end{eqnarray}
where \begin{eqnarray} V(\phi) = \frac{U}{F^{2}} = \frac{FR -f}{2\kappa^2F^2}.   \label{g10}\end{eqnarray}
On varying the action (\ref{g9}) w.r.t. $\phi$ in the absence of matter (relativistic and non-relativistic, both), we get
\begin{eqnarray} \frac{d^{2}\phi}{d\tilde{t}^{2}} + 3\tilde{H}\frac{d\phi}{d\tilde{t}} + V_{,\phi} = 0. \label{g11}\end{eqnarray}
with $V_{,\phi}$  implying the usual derivative w.r.t. $\phi$.  The energy density and pressure of the above  homogeneous scalar field, respectively, are
\begin{eqnarray} \rho = \frac{1}{2}\dot{\phi}^{2} + V(\phi) ;  p = \frac{1}{2}\dot{\phi}^{2} - V(\phi), \label{g12}\end{eqnarray}
while  the scalar field equation of motion is given by (\ref{g11}).

Tolman described a cyclic universe with progressively larger cycles, assuming the presence of a viscous fluid with pressure
\begin{eqnarray} p = p_{0} - 3\zeta H, \label{g13}\end{eqnarray}where $p_{0}$ is the equilibrium pressure and $\zeta$ is the coefficient of bulk viscosity\cite{a14}. It is clear from equation (\ref{g13}) that $p<p_{0}$ during expansion $(H>0)$ whereas $p>p_{0}$ during contraction. This asymmetry during the expanding and contracting phases results in the growth of both energy and entropy. This increase in entropy makes the amplitude of successive expansion cycles larger leading to a arrow of time.

In our discussion of $f(R)$ gravity models, the term $3\tilde{H}\frac{d\phi}{d\tilde {t}}$ in (\ref{g11})  behaves like friction and damps the motion of the scalar field when the universe $(H>0)$. In a contracting universe, however, this  term  behaves like anti-friction and accelerates the motion of the scalar field. A scalar field with the potential $V = M^{2}\phi^{2}$ gives $p \simeq -\rho$  when $H>0$ and $p\simeq\rho$ when $H<0$.
These results are in conformity with  those of Tolman.

Further, we derive different potentials in  all the phases of the Universe. In radiation dominated phase we have $f(R) = \alpha R,$ therefore, the  potential given by equation (\ref{g10})is  $V(\phi)=0$ for this phase. Similarly,  for matter dominated phase we have $f(R) = \beta R$  and $V(\phi) =0$. For accelerated expansion phase,  the form of the Lagrangian is given by including  $f(R)=\gamma R^2$ and the potential for this phase is $V(\phi) =\frac{1}{8\gamma \kappa^2}$.

The scalar field given by equation (\ref{g8})  gives $\kappa\phi =\sqrt{\frac{3}{2}}\ln \alpha $ for matter dominated phase and $\kappa\phi =\sqrt{\frac{3}{2}}\ln \beta $ for radiation dominated phase. For accelerated expansion phase
\begin{eqnarray} \kappa\phi = \sqrt{\frac{3}{2}}\ln (2\gamma R)  \label{g14} \end{eqnarray}
The general solution of the equation (\ref{g11})   for the  potential $V(\phi)=\frac{1}{2}M^{2}\phi^{2}$ is given by
\begin{eqnarray}\phi=\frac{\phi_{0}}{2}\exp(-\Psi t)\left(1+\frac{\Psi}{\sqrt{\Psi^2-M^2}}\right)\exp(\sqrt{\Psi^2 -M^2})t+ \nonumber\\ \frac{\phi_{0}}{2}\exp(-\Psi t) \left(1-\frac{\Psi}{\sqrt{\Psi^2-M^2}}\right)\exp(-\sqrt{\Psi^2-M^2})t \nonumber \\
 \label{g15} \end{eqnarray}
where $\phi_{0}$  is the maximum value of scalar field $\phi$ at $t=0$  ($t$  being the time in the Einstein frame  henceforth), $\Psi=\frac{3\tilde{H}}{2}$,  and $M$ is the mass of the scalar field.
In this solution three cases arises depending on the value of  $\sqrt{\Psi^2 - M^2}$ as discussed below.

\subsection{Case (i): $M^2>\Psi^2$}
\begin{figure}[h]
\centering  \begin{center} \end{center}
\includegraphics[width=0.50\textwidth,origin=c,angle=0]{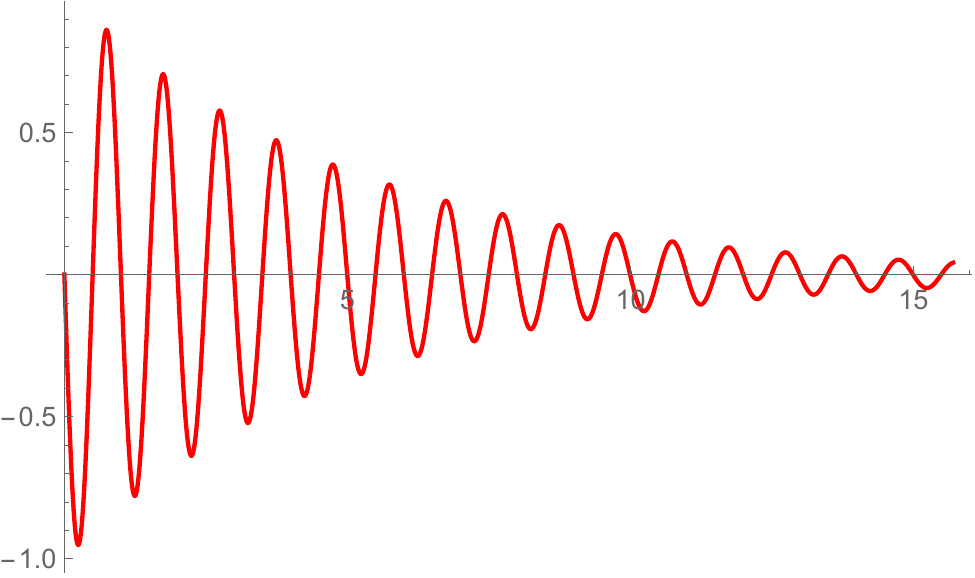}
% "\includegraphics" is very powerful; the graphicx package is already loaded
\caption{\label{fig:p5}Plot for variation of scalar field $\phi$ along $Y-$axis with time  $t$ along $X-$axis in case $M^2>\Psi^2$ . Here, we have taken $\Psi=\frac{1}{5}$, $\frac{\phi_{0}M}{\omega}=1 $,  $\omega=2\pi$ and $\theta=\pi$. }\label{f7}
\end{figure}

In this case the solution is given by
\begin{eqnarray} \phi = \frac{\phi_{0}M}{\omega}\exp(-\Psi t)\sin(\omega t + \theta)    \label{16} \end{eqnarray}
where $\omega = \sqrt{M^2 - \Psi^2}$ and $\theta$ is the phase angle.
If we take $\Psi$ as  a positive constant, then figure (\ref{f7}) shows the nature of scalar field in this case.
These oscillations are damped harmonic oscillations.  In the perturbations due to  local gravity, a very large field mass corresponds to smaller  deviations from the standard $\Lambda$CDM model.
%-------------------------------------------------------------------------------------------
%---------------------------------------------------------------------------------------------

\subsection{Case (ii): $M^2<\Psi^2$}
\begin{figure}[h]
\centering  \begin{center} \end{center}
\includegraphics[width=0.50\textwidth,origin=c,angle=0]{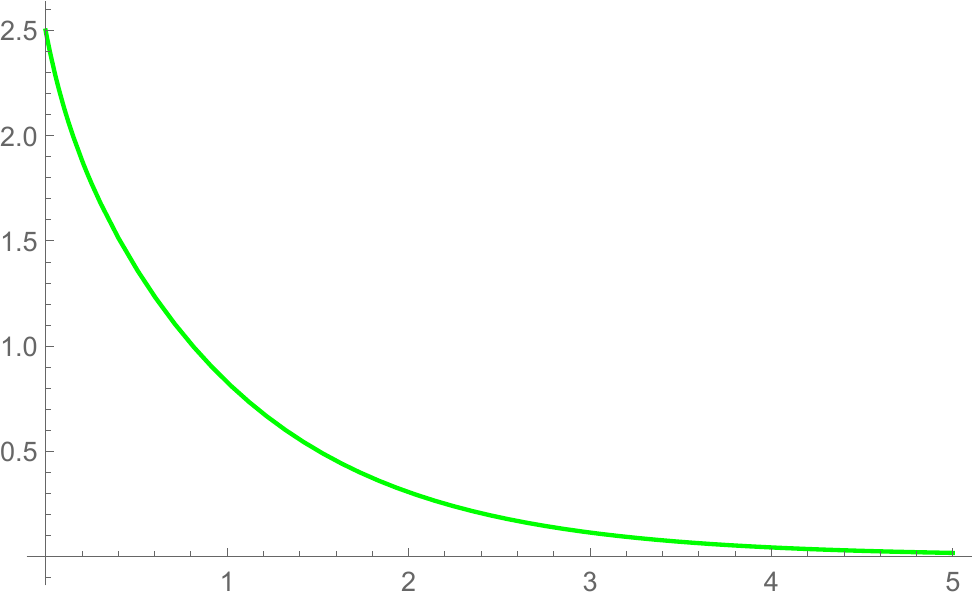}
% "\includegraphics" is very powerful; the graphicx package is already loaded
\caption{\label{fig:p5}Plot for variation of the scalar field $\phi$ along $Y-$axis with time  $t$ along $X-$axis in case $M^2<\Psi^2$. The values of the parameters are as  $\Psi=5$, $M=3$, and $\phi_{0}=2$. }\label{f8}
\end{figure}

The general solution of field equation is given by
\begin{eqnarray} \phi=\frac{\phi_{0}}{2}\exp(-\Psi t)\left(1+\frac{\Psi}{\sqrt{\Psi^2-M^2}}\right)\exp(\sqrt{\Psi^2 -M^2})t+ \nonumber\\ \frac{\phi_{0}}{2}\exp(-\Psi t) \left(1-\frac{\Psi}{\sqrt{\Psi^2-M^2}}\right)\exp(-\sqrt{\Psi^2-M^2})t  \nonumber \\    \label{17} \end{eqnarray}
In this case $(\Psi^2-M^2)$ is a positive quantity and there is an exponential term with negative power. So, the field dies off exponentially with time. There is no oscillation and the motion become  over-damped.  Figure (\ref{f8}) shows the behaviour of the scalar field $\phi$ with time $t$.
%-------------------------------------------------------------------------------------------
%---------------------------------------------------------------------------------------------

\subsection{Case (iii):  $M^2=\Psi^2$}
\begin{figure}[h]
\centering  \begin{center} \end{center}
\includegraphics[width=0.50\textwidth,origin=c,angle=0]{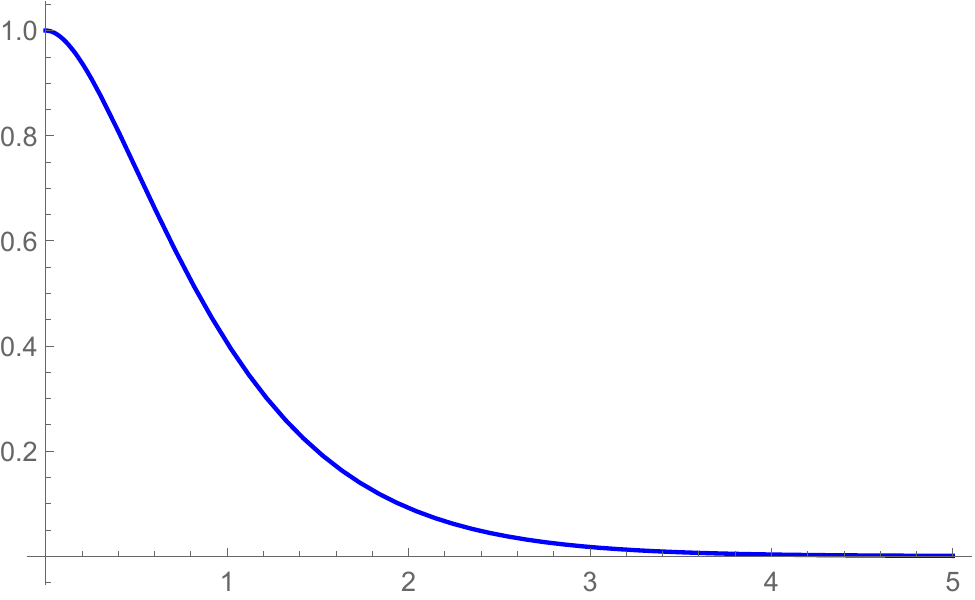}
% "\includegraphics" is very powerful; the graphicx package is already loaded
\caption{\label{fig:p5}Plot for variation of scalar field $\phi$  along $Y-$axis with time  $t$ along $X-$axis in case $M^2=\Psi^2$. Here, we have taken $\Psi=2 $ and $\phi_{0}=1$. }\label{f9}
\end{figure}

This is a special case, appearing as  the critical damping of the scalar field. The  solution is given as

\begin{eqnarray}  \phi = \phi_{0}\exp(-\Psi t)(1+\Psi t)   \label{18} \end{eqnarray}
indicating an aperiodic damping. Figure (\ref{f9}) shows the nature of the scalar field  $\phi$  with the positive $\Psi$.
%-------------------------------------------------------------------------------------------
%---------------------------------------------------------------------------------------------

It is clear that in  all the  above  cases we get a damping  either periodic or  aperiodic that owes its relation to  $\Psi = \frac{3\tilde{H}}{2}$ and mass $M$ of the scalar field $\phi$. When $\tilde{H}$ is  positive and constant, $\Psi$ damps the  scalar field. When $\Psi$ is negative and constant, the motion of the scalar field accelerates in all the three cases.

  We also find that  the scalar field is not symmetric under the reversal of time i.e. $(t\leftrightarrow-t)$ for positive and negative $\Psi=\frac{3\tilde{H}}{2}$. This dissipation  of the scalar field indicates  a crucial  asymmetry in time in form of its arrow from past to future.

\section{\label{10}Conclusion}
 While describing the $f(R)$ models of  modified gravity, we have studied  the properties and stability of the  fixed points of the dynamical system against the time dependent perturbations in a smooth spatial background.  We discussed  the  role of  radiation in our analysis and compared it  with the case without radiation. It is found  that the  nature of the fixed points with radiation remains  unaltered  as that without radiation (except  that with radiation we have the emergence of  an extra eigenvalue for each point). (Of course, the future discussion would bring out an analysis of the fixed points against spatial  perturbations as well).
  We have determined the forms  of $f(R)$ for different phases of the universe, over radiation, matter and acceleration  dominated eras,  by   using the necessary  conditions for the  phase space analysis to reach eventually  at  a cosmologically viable model. The scale factor $a(t)$, the Hubble parameter  $H(t)$, Ricci scalar $R$  have been determined for these phases, with a view that their  ordering over the entire evolution of the universe  may explain the emergence of an arrow of time, more comprehensively  in  a future study.  While  these model  parameters are found to be consistent with  $\Lambda$CDM model,  the crucial issue is that the scalar field $\phi$,  that owes its origin exclusively to $f(R)$ gravity,  may be invoked to explain the arrow of time based on its explicit asymmetry  on a classical level. While the scalar-tensor theories may have the  forms of potentials  matching with $f(R)$,  the results obtained on the  stability conditions  and their ordering  extending through the  overall  history  may not be reproduced in such theories. We would further explore this fundamental aspect of the nature of time  and its observational viability within the  modified gravity sector.

\begin{acknowledgements}
MMV thanks Edward Kolb at the Kavli Institute of Cosmological  Physics, the University of Chicago, USA for hosting the visit, and Varun Sahni at the  Inter-University Centre for Astronomy and Astrophysics (IUCAA), Pune, India for  discussion and  facilities. Both authors thank  IUCAA for the  support under the associateship programme.
\end{acknowledgements}

\end{document}